\documentclass[showpacs,preprintnumbers,superscriptaddress,onecolumn]{revtex4}
\usepackage{amsmath}
\usepackage{graphicx}
\usepackage{dcolumn}
\usepackage{bm}

\input{tcilatex}

\begin{document}

\preprint{APS/123-QED}
\title{One-Photon and Two-Photon Double-Slit Interferences in Spontaneous
and Stimulated Parametric Down-Conversions}
\author{De-Zhong Cao}
\author{Zhuan Li}
\author{Yan-Hua Zhai}
\affiliation{Department of Physics, Applied Optics Beijing Area Major Laboratory, Beijing
Normal University, Beijing 100875, China }
\author{Kaige Wang\thanks{%
the corresponding author}}
\email{wangkg@bnu.edu.cn}
\affiliation{CCAST (World Laboratory), P. O. Box 8730, Beijing 100080, China\\
}
\affiliation{Department of Physics, Applied Optics Beijing Area Major Laboratory, Beijing
Normal University, Beijing 100875, China }
\date{\today }

\begin{abstract}
We theoretically discuss one-photon and two-photon double-slit interferences
for spontaneous and stimulated parametric down-conversions. We show that the
two-photon sub-wavelength interference can exist in a general spontaneous
parametric down-conversion (SPDC) for both type I and type II crystals. We
propose an alternative way to observe sub-wavelength interference by a
joint-intensity measurement which occurs for only type I crystal in a higher
gain of SPDC. When a signal beam injects into the crystal, it may create two
interference patterns by two stimulated down-converted beams, showing no
sub-wavelength interference effect.
\end{abstract}

\maketitle

\section{Introduction}

The Young's double-slit interference experiment is one of the powerful ways
to exhibit the nature of optical field, both classical and nonclassical
coherence effects. In the recent years, an interesting subject is devoted to
the study of two-photon double-slit interference in the process of
spontaneous parametric down-conversion (SPDC).\cite{shih1}-\cite{gigi} Since
in this process a pair of converted beams created by a pump beam is in
entanglement, the two-photon double-slit interference may show some peculiar
phenomena such as the quantum sub-wavelength lithography and the ghost
interference. For the former, both signal and idler beams are set together
to pass through a double-slit,\cite{fon1},\cite{fon2},\cite{sal}-\cite{ab2},%
\cite{shi},\cite{bri} and for the latter, the double-slit is placed on the
path for only one beam.\cite{shih1}-\cite{bar},\cite{fon3},\cite{wal},\cite%
{gigi} The original idea of the quantum lithography comes from the reduction
of de Broglie wavelength for combining two massive particles. In optical
system, the sub-wavelength interference occurs for a biphoton state.\cite%
{yama} In addition to the two-photon double-slit interference, the quantum
lithography can be carried out in a Mach-Zehnder interferometer.\cite{boto}-%
\cite{eda} Due to the fact that this effect can overcome the Rayleigh
diffraction limit, it may have prospective application in photo-lithography
technology.

In most of the theoretical analyses, the sub-wavelength interference is
explained by a two-photon entangled state which can be acquired in the SPDC
with very low gain. Nevertheless, it is the obstacle in practical
application due to the lower power. Therefore, the exploration of these
quantum effects in macroscopic regime makes sense.\cite{gigi},\cite{na} In
this paper, we study one-photon and two-photon double-slit interferences in
both spontaneous and stimulated parametric down-conversions. We focus on the
case in which a double-slit is inserted on the paths for both signal and
idler beams. We find that the sub-wavelength lithography can occur at very
high gain of SPDC with substantial visibility. The discussion covers both
type I and type II crystals which exhibit different behaviors in two-photon
interference. In the stimulated process, the amplified beam cannot perform
quantum lithography but create rich interference patterns. The paper is
organized as follows: in Sec. II we briefly review double-slit interferences
for a coherent state and a two-photon state. In Sec. III we cite several
formula as review of the optical parametric down-conversion process. We
analyze two-photon double-slit interference in Secs. IV and V for the
spontaneous and stimulated processes, respectively. The final section VI is
the conclusion and discussion.

\section{Double-slit Interferences for a Coherent State and a Two-Photon
State}

We consider the scheme of Young's double-slit experiment as shown in Fig. 1.
The double-slit function is defined by 
\begin{equation}
T(x)=\func{rect}(\frac{x-d/2}{b})+\func{rect}(\frac{x+d/2}{b})\text{,}
\label{1}
\end{equation}%
where $d$ is the distance between the centers of two slits and $b$ is the
width of each slit. In Fig. 1, both the double-slit and the detection screen
are placed at the two focus planes of a lens. By ignoring the thickness of
the double-slit, the transverse envelope operators of the input field $%
e(x,t) $ and the output field $e^{\prime }(x,t)$ of the double-slit are
related as 
\begin{equation}
e^{\prime }(x,t)=T(x)e(x,t)+[1-T(x)]e_{vac}(x,t)\text{,}  \label{2}
\end{equation}%
where the vacuum field operator $e_{vac}(x,t)$ is introduced for the sake of 
$e^{\prime }(x,t)$ satisfying the bosonic commutation relation. Since the
vacuum field has no contribution to the normal-order correlation, it can be
neglected in the calculations below.

In the paraxial approximation, the field $r(x,t)$ in the detection plane $%
P_{2}$ is expressed by the Fourier transform of the lens 
\begin{equation}
r(x,t)=\sqrt{\frac{k}{2\pi f}}\int e^{\prime }(x^{\prime },t)\exp [-%
{\normalsize i}\frac{k}{f}x^{\prime }x]dx^{\prime }.  \label{3}
\end{equation}%
By substituting Eq. (\ref{2}) into Eq. (\ref{3}), one obtains 
\begin{equation}
r(x,t)=\frac{1}{2\pi }\sqrt{\frac{k}{f}}\diint \widetilde{T}(\frac{kx}{f}-q)%
\widetilde{e}(q,\Omega )\exp [-{\normalsize i}\Omega t]dqd\Omega \text{,}
\label{4}
\end{equation}%
where 
\begin{equation}
\widetilde{T}(q)=\frac{1}{\sqrt{2\pi }}\int T(x)e^{-iqx}dx=\frac{2b}{\sqrt{%
2\pi }}\func{sinc}(qb/2)\cos (qd/2)  \label{5}
\end{equation}%
is the Fourier transform of the double-slit function $T(x)$, and $\widetilde{%
e}(q,\Omega )$\ is the Fourier transform\ of $e(x,t)$ for both the spatial
and temporal variables.

First, we consider the input field to be a stationary, monochromatic plane
wave in a coherent state 
\begin{equation}
\left\langle e(x,t)\right\rangle =A\text{,}  \label{6}
\end{equation}%
where $A$ is a constant. It has%
\begin{equation}
\left\langle \widetilde{e}(q,\Omega )\right\rangle =2\pi A\delta (q)\delta
(\Omega )\text{.}  \label{7}
\end{equation}%
\bigskip In the detection plane, the first-order correlation is calculated as%
\begin{equation}
G^{(1)}(x_{1},x_{2},t)\equiv \left\langle r^{\dagger
}(x_{1},t)r(x_{2},t)\right\rangle =\frac{kA^{2}}{f}\widetilde{T}^{\ast }(%
\frac{kx_{1}}{f})\widetilde{T}(\frac{kx_{2}}{f}).  \label{g1}
\end{equation}%
Hence, the intensity distribution in the detection plane is written as 
\begin{equation}
G^{(1)}(x,x,t)\equiv \left\langle r^{\dagger }(x,t)r(x,t)\right\rangle =%
\frac{kA^{2}}{f}\widetilde{T}^{2}(\frac{kx}{f})=I_{0}\func{sinc}^{2}(\frac{%
\pi bx}{\lambda f})\cos ^{2}(\frac{\pi dx}{\lambda f})  \label{8}
\end{equation}%
where $I_{0}=\frac{2kb^{2}A^{2}}{\pi f}$ and $\lambda =2\pi /k$.\ Note that $%
\widetilde{T}(x)$ is a real function. Equation (\ref{8}) represents an
interference fringe with the interval $\lambda f/d$ in the range $\lambda f/b
$, as shown in Fig. 2.

Similarly, the second-order correlation function can be obtained as 
\begin{eqnarray}
G^{(2)}(x_{1},x_{2},t) &=&\left\langle r^{\dagger }(x_{1},t)r^{\dagger
}(x_{2},t)r(x_{2},t)r(x_{1},t)\right\rangle  \notag \\
&=&\frac{k^{2}A^{4}}{f^{2}}\widetilde{T}^{2}(\frac{kx_{1}}{f})\widetilde{T}%
^{2}(\frac{kx_{2}}{f}).  \label{9}
\end{eqnarray}%
According to the theory of field coherence, the separability of spatial
variables in the first-order and the second-order correlation functions
verifies the perfect coherence of the field. Since the field operators at
different positions are commutable, the second-order correlation of the
field is in fact the spatial intensity correlation\ and it can be observed
by a coincidence measurement as shown in Fig. 1a. The spatial patterns
related to $G^{(1)}(x,x,t)$ and $G^{(2)}(x_{1},x_{2},t)$ are called the
one-photon and the two-photon interferences, respectively. According to Eq. (%
\ref{9}), in the coincidence measurement, if we scan one detector by fixing
another, the same interference fringe as the one-photon interference can be
observed. Now, we introduce two other ways of observations for the
two-photon double-slit interference. One is the spatial
intensity-correlation measurement by scanning two detectors synchronously at
a pair of symmetric positions, $x_{1}=-x_{2}=x$. The other one is the
two-photon intensity measurement by using a two-photon detector which
generates an photo-electron by absorbing two photons. Applying these two
observations to Eq. (\ref{9}), one has%
\begin{equation}
G^{(2)}(x,x,t)=G^{(2)}(x,-x,t)=I_{0}^{2}\func{sinc}^{4}(\frac{\pi bx}{%
\lambda f})\cos ^{4}(\frac{\pi dx}{\lambda f}).  \label{10}
\end{equation}%
In Fig. 2, we plot $G^{(2)}(x,x,t)$ ($G^{(2)}(x,-x,t)$) in comparison with $%
G^{(1)}(x,t)$ for the coherent beam. It shows that the two interference
patterns are alike.

The above discussion on the coherent state is analogous to the classical
field. Then, we consider a two-photon state as input, which is the quantum
state without classical analogue. A general two-photon state can be written
as%
\begin{equation}
|\psi \rangle =\dint dq_{s}dq_{i}C(q_{s},q_{i})a_{s}^{\dagger
}(q_{s})a_{i}^{\dagger }(q_{i})|0\rangle ,  \label{t1}
\end{equation}%
where $a_{s}^{\dagger }$ and $a_{i}^{\dagger }$ are the creation operators
for $s$\ and $i$ photons which are assumed to be distinguishable. $q_{s}$
and $q_{i}$ are the transverse wave-vectors. When the input field is
stationary, Eq. (\ref{4}) can be simplified as%
\begin{equation}
r(x)=\sqrt{\frac{k}{2\pi f}}\dint \widetilde{T}(\frac{kx}{f}-q)\widetilde{e}%
(q)dq.  \label{t2}
\end{equation}%
By using Eq. (\ref{t2}), the first-order correlation functions for $s$%
-photon and $i$-photon in the detection plane are obtained as 
\begin{subequations}
\label{t3}
\begin{eqnarray}
G_{s}^{(1)}(x_{1},x_{2}) &=&\frac{k}{2\pi f}\int dqdq_{1}dq_{2}C^{\ast
}(q_{1},q)C(q_{2},q)\widetilde{T}^{\ast }(\frac{kx_{1}}{f}-q_{1})\widetilde{T%
}(\frac{kx_{2}}{f}-q_{2}),  \label{t3a} \\
G_{i}^{(1)}(x_{1},x_{2}) &=&\frac{k}{2\pi f}\int dqdq_{1}dq_{2}C^{\ast
}(q,q_{1})C(q,q_{2})\widetilde{T}^{\ast }(\frac{kx_{1}}{f}-q_{1})\widetilde{T%
}(\frac{kx_{2}}{f}-q_{2}),  \label{t3b}
\end{eqnarray}%
respectively.

For two fields case, the second-order correlation function is defined by 
\end{subequations}
\begin{equation}
G^{(2)}(x_{1},x_{2},t)=\left\langle r_{i}^{\dagger }(x_{1},t)r_{s}^{\dagger
}(x_{2},t)r_{s}(x_{2},t)r_{i}(x_{1},t)\right\rangle ,  \label{t4}
\end{equation}%
which describes the coincidence probability of $s$-photon at position $x_{2}$
and $i$-photon at position $x_{1}$. In the case of $x_{1}=x_{2}=x$, it
describes a two-photon intensity distribution. First, for the two-photon
state (\ref{t1}), we calculate the two-photon wavepacket in the detection
plane%
\begin{equation}
\langle 0|r_{s}(x_{2})r_{i}(x_{1})|\psi \rangle =\frac{k}{2\pi f}\int
dq_{s}dq_{i}C(q_{s},q_{i})\widetilde{T}(\frac{kx_{2}}{f}-q_{s})\widetilde{T}(%
\frac{kx_{1}}{f}-q_{i}).  \label{t5}
\end{equation}%
Then the second-order correlation is obtained as%
\begin{equation}
G^{(2)}(x_{1},x_{2})=|\langle 0|r_{s}(x_{2})r_{i}(x_{1})|\psi \rangle |^{2}.
\label{t6}
\end{equation}

We discuss two extreme cases: two photons are independent and perfectly
entangled in the transverse wave-vector. In the unentangled case, $%
C(q_{s},q_{i})=C$\bigskip $_{s}(q_{s})C_{i}(q_{i})$, the first- and
second-order correlations are written as 
\begin{subequations}
\label{t7}
\begin{eqnarray}
G_{m}^{(1)}(x_{1},x_{2}) &=&\frac{k}{2\pi f}\int dqC_{m}^{\ast }(q)%
\widetilde{T}^{\ast }(\frac{kx_{1}}{f}-q)\int dqC_{m}(q)\widetilde{T}(\frac{%
kx_{2}}{f}-q),\qquad (m=s,i),  \label{t7a} \\
G^{(2)}(x_{1},x_{2}) &=&\left\vert \frac{k}{2\pi f}\int dqC_{s}(q)\widetilde{%
T}(\frac{kx_{2}}{f}-q)\int dqC_{i}(q)\widetilde{T}(\frac{kx_{1}}{f}%
-q)\right\vert ^{2}  \notag \\
&=&G_{i}^{(1)}(x_{1},x_{1})G_{s}^{(1)}(x_{2},x_{2}).  \label{t7b}
\end{eqnarray}%
Equations (\ref{t7}) show two separabilities. The one is the separability of
positions in both the first- and second-order correlation functions. The
other is that the second-order correlation is factorized to two one-photon
intensity distribution. That is, the two-photon interference consists of two
individual one-photon interferences, verifying the Dirac's statement: "Each
photon interferes only with itself. Interference between two different
photons never occurs." However, these features mean the perfect first- and
second-order coherencies for the two independent photons.

In the opposite extreme, the perfect entanglement in wavevector, $%
q_{s}+q_{i}=0$, is hold in state (\ref{t1}). For simplicity, we assume $%
C(q_{s},q_{i})\rightarrow \delta (q_{s}+q_{i})$, and the first- and
second-order correlations are written as 
\end{subequations}
\begin{subequations}
\label{t8}
\begin{eqnarray}
G_{s}^{(1)}(x_{1},x_{2}) &=&G_{i}^{(1)}(x_{1},x_{2})=\frac{k}{2\pi f}\int dq%
\widetilde{T}^{\ast }(\frac{kx_{1}}{f}+q)\widetilde{T}(\frac{kx_{2}}{f}+q)=%
\frac{k}{f\sqrt{2\pi }}\widetilde{T}[\frac{k}{f}(x_{2}-x_{1})],  \label{t8a}
\\
G^{(2)}(x_{1},x_{2}) &=&\left\vert \frac{k}{2\pi f}\int dq\widetilde{T}(%
\frac{kx_{2}}{f}+q)\widetilde{T}(\frac{kx_{1}}{f}-q)\right\vert ^{2}=\frac{%
k^{2}}{2\pi f^{2}}\widetilde{T}^{2}[\frac{k}{f}(x_{1}+x_{2})],  \label{t8b}
\end{eqnarray}%
where we use the integrals 
\end{subequations}
\begin{eqnarray}
\int dq\widetilde{T}^{\ast }(\frac{kx_{1}}{f}\pm q)\widetilde{T}(\frac{kx_{2}%
}{f}\pm q) &=&\frac{1}{2\pi }\iiint dqdx_{1}^{\prime }dx_{2}^{\prime
}T(x_{1}^{\prime })T(x_{2}^{\prime })e^{i(\frac{kx_{1}}{f}\pm
q)x_{1}^{\prime }-i(\frac{kx_{2}}{f}\pm q)x_{2}^{\prime }}  \label{t9} \\
&=&\diint dx_{1}^{\prime }dx_{2}^{\prime }T(x_{1}^{\prime })T(x_{2}^{\prime
})\delta (x_{1}^{\prime }-x_{2}^{\prime })e^{i\frac{kx_{1}}{f}x_{1}^{\prime
}-i\frac{kx_{2}}{f}x_{2}^{\prime }}  \notag \\
&=&\dint dx_{1}^{\prime }T(x_{1}^{\prime })T(x_{1}^{\prime })e^{i\frac{k}{f}%
(x_{1}-x_{2})x_{1}^{\prime }}=\sqrt{2\pi }\widetilde{T}[\frac{k}{f}%
(x_{2}-x_{1})],  \notag
\end{eqnarray}%
and%
\begin{equation}
\int dq\widetilde{T}^{\ast }(\frac{kx_{1}}{f}\pm q)\widetilde{T}(\frac{kx_{2}%
}{f}\mp q)=\sqrt{2\pi }\widetilde{T}[\frac{k}{f}(x_{2}+x_{1})].  \label{t10}
\end{equation}%
Note that $T^{2}(x)=T(x)$. Equations (\ref{t8}) show that for the
wavevector-entangled two-photon state the first- and second-order
correlation functions in double-slit interference have position-correlation
which results in decoherence. In the measurement, letting $x_{1}=x_{2}$ in
Eq. (\ref{t8}), one obtains 
\begin{subequations}
\label{t11}
\begin{eqnarray}
G_{s}^{(1)}(x,x) &=&G_{i}^{(1)}(x,x)=\frac{k}{f\sqrt{2\pi }}\widetilde{T}[0],
\label{t11a} \\
G^{(2)}(x,x) &=&\frac{k^{2}}{2\pi f^{2}}\widetilde{T}^{2}[\frac{k}{f}(2x)].
\label{t11b}
\end{eqnarray}%
Therefore, the one-photon double-slit interference disappears completely and
the two-photon double-slit interference shows a sub-wavelength property
since it has 
\end{subequations}
\begin{equation}
\widetilde{T}^{2}(\frac{k}{f}2x)\propto \func{sinc}^{2}[\frac{\pi bx}{%
(\lambda /2)f}]\cos ^{2}[\frac{\pi dx}{(\lambda /2)f}].  \label{31}
\end{equation}%
The fringe is the same as the ordinary double-slit interference with the
half of the wavelength. The above analysis explains the complementarity of
coherence and entanglement.\cite{sal},\cite{ab1},\cite{ab2} We emphasize
that the discussion is also true for the case when $s$\ and $i$ photons can
be indistinguishable.

\section{The Basic Formula in the Optical Parametric Down-Conversion}

In the optical parametric down-conversion, in which a plane-wave pump field
of frequency $\omega _{p}$ activates a $\chi ^{(2)}$ nonlinear crystal, the
basic unitary transformation is described by\cite{gigi}, \cite{12}-\cite{14} 
\begin{equation}
\widetilde{e}_{m}(q,\Omega )=U_{m}(q,\Omega )\widetilde{a}_{m}(q,\Omega
)+V_{m}(q,\Omega )\widetilde{a}_{n}^{\dagger }(-q,-\Omega )\qquad (m\neq
n=s,i),  \label{11}
\end{equation}%
where $\widetilde{e}_{m}(q,\Omega )$ and $\widetilde{a}_{m}(q,\Omega )$ are
the output and input field operators, respectively. $q$ is the transverse
wavevector and $\Omega $\ is the frequency deviation from the carrier
frequency. The transfer coefficients\ $U_{m}(q,\Omega )$ and $V_{m}(q,\Omega
)$ are given by\cite{14} 
\begin{equation}
U_{s}(q,\Omega )=\Theta _{s}(q,\Omega )[\cosh \Gamma (q,\Omega )+i\frac{%
\Delta (q,\Omega )}{2\Gamma (q,\Omega )}\sinh \Gamma (q,\Omega )],
\label{14}
\end{equation}%
\begin{equation}
V_{s}(q,\Omega )=\Theta _{s}(q,\Omega )\frac{g}{\Gamma (q,\Omega )}\sinh
\Gamma (q,\Omega ),  \label{15}
\end{equation}%
\begin{equation}
U_{i}(q,\Omega )=\Theta _{i}(q,\Omega )[\cosh \Gamma (-q,-\Omega )+i\frac{%
\Delta (-q,-\Omega )}{2\Gamma (-q,-\Omega )}\sinh \Gamma (-q,-\Omega )],
\label{16}
\end{equation}%
\begin{equation}
V_{i}(q,\Omega )=\Theta _{i}(q,\Omega )\frac{g}{\Gamma (-q,-\Omega )}\sinh
\Gamma (-q,-\Omega ),  \label{17}
\end{equation}%
where 
\begin{equation}
\Theta _{m}(q,\Omega )=e^{i[k_{mz}(q,\Omega )-k_{nz}(-q,-\Omega
)-2k_{m}+k_{p}]l_{c}/2},\qquad (m\neq n=s,i),  \label{18}
\end{equation}%
\begin{equation}
\Gamma (q,\Omega )=\sqrt{g^{2}-\Delta ^{2}(q,\Omega )/4},  \label{19}
\end{equation}%
\begin{equation}
\Delta (q,\Omega )=[k_{sz}(q,\Omega )+k_{iz}(-q,-\Omega )-k_{p}]l_{c},
\label{20}
\end{equation}%
\begin{equation}
\Delta _{0}=(k_{s}+k_{i}-k_{p})l_{c}.  \label{21}
\end{equation}%
$g$ is the coupling strength and $l_{c}$ is the length of crystal. $\Delta
_{0}$ is the collinear phase mismatching of the central frequency components
which correspond to the wave-numbers $k_{j}$ ($j=s,i,p$). For simplicity, we
assume that two down-converted beams have the degenerate carrier frequency $%
\omega _{p}/2$. Hence, Eq. (\ref{20}) can be reduced to an even function of
both $q$ and $\Omega $%
\begin{equation}
\Delta (q,\Omega )\approx \Delta _{0}+\Omega ^{2}/\Omega
_{0}^{2}-q^{2}/q_{0}^{2},  \label{20p}
\end{equation}%
where $\Omega _{0}$ and $q_{0}$ are defined as the frequency and
spatial-frequency bandwidths, respectively.

Equations (\ref{11})-(\ref{20p}) describe the SPDC process of type II
crystal but also can be applicable to type I crystal. For the former, two
converted beams are orthogonally polarized, whereas for the latter, they are
degenerate in both polarization and frequency but spatially separated.
Therefore, Eq. (\ref{11}) can describe type I crystal by omitting the
subscripts. As a matter of fact, under the assumption of the carrier
frequency degeneracy, Eqs. (\ref{14}) and (\ref{15}) are the same as Eqs. (%
\ref{16}) and (\ref{17}).

\section{Double-Slit Interference in Spontaneous Parametric Down-Conversion}

In Fig. 1, the two down-converted beams generated from the crystal
illuminate a double-slit and then are detected in the focus plane of the
lens. We designate $a_{m}(x,t)$, $e_{m}(x,t)$,$\ e_{m}^{\prime }(x,t)$, $%
r_{m}(x,t)$ the slowly varying field operators for the input surface $P_{in}$
and the output surface $P_{out}$ of the crystal and the output plane of the
double-slit $P_{1}$ and the detection plane $P_{2}$, respectively.
Substituting Eq. (\ref{11}) into Eq. (\ref{4}), we may calculate the first-
and the second-order correlations for the field in the detection plane $%
P_{2} $.

In this section, we consider the case of the spontaneous parametric
down-conversion in which the input fields are in the vacuum state. The
first-order correlations for two beams are obtained to be 
\begin{eqnarray}
G_{m}^{(1)}(x_{1},x_{2}) &=&M_{m}(x_{1},x_{2})\equiv \langle
0|r_{m}^{\dagger }(x_{1},t)r_{m}(x_{2},t)|0\rangle \qquad \qquad \qquad
\qquad \qquad \qquad \qquad \qquad (m=s,i)  \label{22} \\
&=&\frac{k/f}{(2\pi )^{2}}\iint dqd\Omega \quad \left\vert V_{m}(q,\Omega
)\right\vert ^{2}\widetilde{T}^{\ast }(\frac{kx_{1}}{f}-q)\widetilde{T}(%
\frac{kx_{2}}{f}-q).  \notag
\end{eqnarray}%
$G_{m}^{(1)}(x,x)$ illustrates the intensity distribution for beam $m$ in
the detection plane. Then we consider the second-order correlation function
defined by Eq. (\ref{t4}), which now describes the intensity correlation
between the signal beam at position $x_{2}$ and the idler beam at position $%
x_{1}$. In the case of $x_{1}=x_{2}=x$, it describes a two-photon intensity
distribution. For type II crystal, the two-photon intensity distribution can
be measured experimentally by the scheme shown in Fig. 1b in which the
coincidence measurement of two orthogonally polarized photons is performed
at the same position $x_{1}=x_{2}=x$. However, for type I crystal, the
subscripts $s$ and $i$ should be omitted in Eq. (\ref{t4}). If there is a
two-photon detector available, one may observe the two-photon intensity
distribution at position $x_{1}=x_{2}=x$. Alternatively, a realistic
detective scheme for type I crystal is shown in Fig. 1a in which the
intensity correlation is measured by two one-photon detectors at the
different positions $x_{1}$ and $x_{2}$.

For the vacuum state set in Eq. (\ref{t4}), by using Eqs. (\ref{4}) and (\ref%
{11}), the second-order correlation is calculated as 
\begin{equation}
G^{(2)}(x_{1},x_{2})=M_{i}(x_{1},x_{1})M_{s}(x_{2},x_{2})+\left\vert
N_{is}(x_{1},x_{2})\right\vert ^{2}+\delta _{is}\left\vert
M(x_{1},x_{2})\right\vert ^{2},  \label{g2}
\end{equation}%
where $\delta _{is}$ is 1 for type I and 0 for type II crystal. $%
M_{m}(x_{1},x_{2})$ is given by Eq. (\ref{22}), and

\begin{equation}
N_{mn}(x_{{\normalsize 1}},x_{{\normalsize 2}})=\frac{k/f}{(2\pi )^{2}}\iint
dqd\Omega \quad V_{m}(q,\Omega )U_{n}(-q,-\Omega )\widetilde{T}(\frac{kx_{%
{\normalsize 1}}}{f}-q)\widetilde{T}(\frac{kx_{{\normalsize 2}}}{f}%
+q).\qquad (m\neq n=s,i)  \label{25}
\end{equation}%
Equation (\ref{g2}) shows that the second-order correlation is related to
the first-order correlation. In Eq. (\ref{g2}), the first term, which is
separable in term of both polarization and position, describes the part of
two-photon interference contributed by two individual single-photon
double-slit processes. However, the second and the third terms describe
two-photon interference effect related to photon entanglement. The
difference between the two types of crystals is devoted by the third term,
that is, the first-order correlation $|M(x_{1},x_{2})|^{2}$.

We note that, since $\Theta _{i}(q,\Omega )\Theta _{s}(-q,-\Omega )=\Theta
_{s}(q,\Omega )\Theta _{i}(-q,-\Omega )=\exp [-i\Delta _{0}]$, it has $%
V_{i}(q,\Omega )U_{s}(-q,-\Omega )=V_{s}(q,\Omega )U_{i}(-q,-\Omega )$, and
hence $N_{is}(x_{{\normalsize 1}},x_{{\normalsize 2}})=N_{si}(x_{%
{\normalsize 1}},x_{{\normalsize 2}})$. This result is obvious because
exchange of the indices $i$ and $s$ makes no difference physically.
Moreover, due to Eq. (\ref{20p}), $\Delta (q,\Omega )$\ is the even function
of $q$ and\ $\Omega $ under the assumption of the frequency degeneracy so
that $\left\vert V_{i}(q,\Omega )\right\vert ^{2}=\left\vert V_{s}(q,\Omega
)\right\vert ^{2}$. Therefore, two first-order correlation functions for the
signal and the idler beams are equal in type II crystal, i.e. $%
M_{i}(x_{1},x_{2})=M_{s}(x_{1},x_{2})$. Nevertheless, we keep the subscripts
in Eq. (\ref{g2}) for a general description of type II crystal in case the
two converted beams have different carrier frequencies.

In order to obtain the analytical result for the integrals,\ we discuss two
bandwidth limits of SPDC process: the broad and the narrow bandwidths. In
the broadband limit, $q_{0}\gg 2\pi /d$, $U_{m}(q,\Omega )$ and $%
V_{m}(q,\Omega )$ are much flatter in comparison with $\widetilde{T}(q)$ and
one can set $U_{m}(q,\Omega )\approx U_{m}(0,\Omega )$ and $V_{m}(q,\Omega
)\approx V_{m}(0,\Omega )$ in the integrals. By taking into account Eqs. (%
\ref{t9}) and (\ref{t10}), Eqs. (\ref{22}) and (\ref{25}) can be rewritten
as 
\begin{subequations}
\begin{equation}
M_{m}(x_{1},x_{2})=\frac{\eta _{m}}{\sqrt{2\pi }}\dint dq\widetilde{T}^{\ast
}(\frac{kx_{1}}{f}-q)\widetilde{T}(\frac{kx_{2}}{f}-q)=\eta _{m}\widetilde{T}%
[\frac{k}{f}(x_{2}-x_{1})],\text{\qquad }(m=s,i),  \label{27a}
\end{equation}%
and

\end{subequations}
\begin{subequations}
\label{27}
\begin{equation}
N_{mn}(x_{1},x_{2})=\xi _{mn}\widetilde{T}[\frac{k}{f}(x_{1}+x_{2})],\text{%
\qquad }(m\neq n=s,i),  \label{27b}
\end{equation}%
respectively, where we define $\eta _{m}=\frac{k/f}{(2\pi )^{3/2}}\int
\left\vert V_{m}(0,\Omega )\right\vert ^{2}d\Omega $ and $\xi _{mn}=\frac{k/f%
}{(2\pi )^{3/2}}\int V_{m}(0,\Omega )U_{n}(0,-\Omega )d\Omega $.

In the broadband limit, which exhibits the maximum entanglement in
transverse wavevector for two converted beam, we see again the
position-correlation in the correlation functions. The first-order
correlation Eq. (\ref{27a}) shows the same position-correlation as Eq. (\ref%
{t8a}) for the two-photon state with the maximum wavevector-entanglement.
This makes the one-photon intensity distribution in the detection plane $%
P_{2}$ homogeneous, $M_{m}(x,x)=\eta _{m}\widetilde{T}(0)$, so that the
one-photon double-slit interference disappears completely.

In this limit, the second-order correlation (\ref{g2}) is obtained to be

\end{subequations}
\begin{equation}
G^{(2)}(x_{1},x_{2})=\eta _{i}\eta _{s}\left\{ \widetilde{T}^{2}(0)+\delta
_{is}\widetilde{T}^{2}[\frac{k}{f}(x_{2}-x_{1})]\right\} +\left\vert \xi
_{is}\right\vert ^{2}\widetilde{T}^{2}[\frac{k}{f}(x_{1}+x_{2})].  \label{28}
\end{equation}%
The first term in \{\} comes from two individual single-photon double-slit
processes which are now homogeneous. The second term in \{\} and the last
term manifest explicitly the position-corrlation. If we fix one detector at
a position and scan another in the coincidence measurement, the interference
fringe observed is the same as the single-photon one. To show the position
correlation in the two-photon interference, we discuss two special
observations, that is, $x_{1}=x_{2}=x$ and $x_{1}=-x_{2}=x$. Equation (\ref%
{28}) is reduced to%
\begin{equation}
G^{(2)}(x,x)=\eta _{i}\eta _{s}(1+\delta _{is})\widetilde{T}%
^{2}(0)+\left\vert \xi _{is}\right\vert ^{2}\widetilde{T}^{2}(\frac{k}{f}2x),
\label{29}
\end{equation}%
and%
\begin{equation}
G^{(2)}(x,-x)=(\eta _{i}\eta _{s}+\left\vert \xi _{is}\right\vert ^{2})%
\widetilde{T}^{2}(0)+\delta _{is}\eta _{i}\eta _{s}\widetilde{T}^{2}(\frac{k%
}{f}2x).  \label{30}
\end{equation}%
The former exhibits a two-photon intensity distribution and the latter
exhibits the intensity correlation at a pair of symmetric positions. Both
Eqs. (\ref{29}) and (\ref{30}) include a term $\widetilde{T}^{2}(\frac{k}{f}%
2x)$ which characterizes a sub-wavelength interference pattern by the factor
of $\lambda /2$\ in comparison with\ the ordinary interference shown by Eq. (%
\ref{8}). Obviously, due to Eq. (\ref{29}), the sub-wavelength interference
for the two-photon intensity distribution can occur in both type I and type
II crystals. However, Eq. (\ref{30}) shows that, for type I crystal, when a
pair of single-photon detectors are placed at a pair of symmetric positions
and moved synchronously in the opposite direction, the sub-wavelength
interference can be also observed. But this effect never happens in type II
crystal.

According to Eqs. (\ref{29}) and (\ref{30}), the visibilities of fringes
designated by $G^{(2)}(x,x)$ and $G^{(2)}(x,-x)$ are calculated to be 
\begin{equation}
\mathcal{V}_{1}=\frac{1}{1+2(1+\delta _{is})\theta },  \label{32}
\end{equation}%
and 
\begin{equation}
\mathcal{V}_{2}=\frac{1}{3+2/\theta },  \label{33}
\end{equation}%
respectively, where $\theta \equiv \eta _{i}\eta _{s}/\left\vert \xi
_{is}\right\vert ^{2}$. Note that $\mathcal{V}_{2}$ makes sense only for
type I crystal, for which $\theta \equiv \eta ^{2}/\left\vert \xi
\right\vert ^{2}$. As the parameter $\theta $ is increased from a very small
quantity, $\mathcal{V}_{1}$ decreases monotonously from unity and $\mathcal{V%
}_{2}$ increases from zero up to 1/3. Since the parameter $\theta $ is
related to the coupling strength $g$ of SPDC, we plot the visibilities as
functions of $g$ in Figs. 3, in which $\mathcal{V}_{1}$s for type II and
type I crystals are indicated by the solid and dashed lines, respectively,
and $\mathcal{V}_{2}$ for type I crystal is indicated by the dotted line. In
a weak coupling of SPDC, which may generate approximately a two-photon
entangled state, the visibilities $\mathcal{V}_{1}$ for both type I and II
crystals reach perfectness. The sub-wavelength interference in the weak SPDC
has been observed experimentally.\cite{fon1},\cite{shih2},\cite{shi},\cite%
{eda} The important fact is that, in Figs. 3, the sub-wavelength
interference can still exist, with a substantial visibility, even in very
high gain of SPDC, in which the beams contain a large amount of photons. On
the other hand, in the strong coupling of type I crystal, the sub-wavelength
interference can occur in the joint-intensity measurement at a pair of
symmetric positions. Equations (\ref{32}) and (\ref{33}) show that, for type
I crystal, two observations of sub-wavelength interferences compete for
visibility. When $\mathcal{V}_{1}$ reaches the perfectness, $\mathcal{V}_{2}$
vanishes. However, in the very high gain of SPDC of type I, two visibilities
are the same as 25\%.

In the opposite limit, we assume that the SPDC has a very narrow bandwidth $%
q_{0}\ll 2\pi /d$.\ Extremely, letting $q_{0}\rightarrow 0$, the transfer
coefficient $V_{m}(q,\Omega )$ ($m=s$,$i$) tends to the delta function 
\begin{equation}
V_{m}(q,\Omega )\rightarrow \ V_{m}(0,\Omega )\delta (q)  \label{34}
\end{equation}%
Equations (\ref{22}) and (\ref{25}) are respectively written as%
\begin{equation}
M_{m}(x_{1},x_{2})=\frac{1}{\sqrt{2\pi }}\eta _{m}\widetilde{T}(\frac{kx_{1}%
}{f})\widetilde{T}^{\star }(\frac{kx_{2}}{f}),  \label{35}
\end{equation}%
and%
\begin{equation}
N_{mn}(x_{1},x_{2})=\frac{1}{\sqrt{2\pi }}\xi _{mn}\widetilde{T}(\frac{kx_{1}%
}{f})\widetilde{T}(\frac{kx_{2}}{f})\text{.}  \label{36}
\end{equation}%
In this limit, the position-correlation disappears completely. The
second-order correlation is then 
\begin{equation}
G^{(2)}(x_{1},x_{2})=\frac{1}{2\pi }[(1+\delta _{is})\eta _{i}\eta _{s}+|\xi
_{is}|^{2}]\widetilde{T}^{2}(\frac{kx_{1}}{f})\widetilde{T}^{2}(\frac{kx_{2}%
}{f}).  \label{37}
\end{equation}%
Therefore, the one-photon intensity distribution $M_{m}(x,x)$ and the
second-order correlation $G^{(2)}(x_{1},x_{2})$ in the plane $P_{2}$ are the
same as the case for the coherent state. In the narrow bandwidth limit,
since each converted beam is monochromatic, the one-photon double-slit
interference occurs with the perfect visibility. On the other hand, two
monochromatic converted beams have no more correlation in the transverse
wavevector so that the position-correlation degrades completely in the
two-photon interference.

We plot the two-photon interference patterns by varying the bandwidth of
SPDC process $q_{0}$\ in Figs. 4, in which Figs. 4a and 4b (4c and 4d) are
for type I (type II) crystal. In Figs. 4a and 4c, a low gain of SPDC, $%
g=(1/2)\log 1.5$ (with the amplification rate 1.5) is taken in two-photon
intensity measurement so that the sub-wavelength interferences with a better
visibility are achieved when the normalized bandwidth $q_{0}b/(2\pi )$ is
increased. Two sets of patterns are very similar with exception of tiny
higher intensity for type I crystal. Figures 4b and 4d show the interference
patterns for the joint-intensity measurement of two one-photon detectors at
a pair symmetric positions. The sub-wavelength interferences can be observed
only for type I crystal when the gain of SPDC is higher, for instance, $%
g=(1/2)\log 10$ (with the amplification rate 10) is taken in the figures.
Though the visibilities are lower, the intensities of the patterns are
getting much higher. This also happens in the case of two-photon intensity
measurement. The three plots of Figs. 4a-4c show that the bandwidth of SPDC
governs two-photon sub-wavelength interference. For a very small bandwidth
of SPDC, the two converted beams are de-entangled in transverse wavevector
so that the nonclassical sub-wavelength interference disappears.

\section{Double-Slit Interference in Stimulated Parametric Down-Conversion}

In the stimulated optical parametric process, a signal beam is injected into
the nonlinear crystal and is then amplified. The nonlinear crystal becomes
an optical parametric amplifier (OPA). We assume inputting a stable
plane-wave beam in a coherent state%
\begin{equation}
\left\langle \widetilde{a}_{s}(q,\Omega )\right\rangle =2\pi A\delta
(q-Q)\delta (\Omega ),  \label{38}
\end{equation}%
where $Q$ designates the transverse wavevector of the input beam deviated
from the normal incidence. For type II crystal, we set the input beam as the
signal which can be identified by the polarization, while the idler beam is
in the vacuum state. For type I crystal, the subscript $s$ in the input beam
(\ref{38}) is omitted. Considering the input beam describing by Eq. (\ref{38}%
) instead of the vacuum state, we calculate the first-order correlation in
the plane $P_{2}$%
\begin{equation}
G_{m}^{(1)}(x_{1},x_{2})=W_{m}^{\ast
}(x_{1},Q)W_{m}(x_{2},Q)+M_{m}(x_{1},x_{2}),\qquad (m=s,i)  \label{39}
\end{equation}%
where 
\begin{subequations}
\label{40}
\begin{eqnarray}
W_{s}(x,Q) &=&A\sqrt{k/f}U_{s}(Q,0)\widetilde{T}(kx/f-Q)  \label{40a} \\
W_{i}(x,Q) &=&A\sqrt{k/f}V_{i}(-Q,0)\widetilde{T}(kx/f+Q)  \label{40b}
\end{eqnarray}%
for type II crystal, and 
\end{subequations}
\begin{equation}
W(x,Q)=A\sqrt{k/f}[U(Q,0)\widetilde{T}(kx/f-Q)+V(-Q,0)\widetilde{T}(kx/f+Q)]
\label{41}
\end{equation}%
for type I crystal. $M_{m}(x_{1},x_{2})$ has been defined by Eq. (\ref{22}).
In Eq. (\ref{39}), the first and the second terms show contributions coming
from the stimulated and the spontaneous processes, respectively, and they
are independent in the one-photon interference pattern. Obviously, the
stimulated part shows the first-order coherence due to the separability of
the spatial variables. Nevertheless, for the input beam is enough strong,
the spontaneous process can be neglected. $G_{m}^{(1)}(x,x)\simeq
|W_{m}(x,Q)|^{2}$ describes an amplified double-slit interference pattern in
comparison with the case without crystal. For type II crystal, the input
signal beam creates two interference patterns: one for the signal beam with
the amplification rate $|U_{s}(Q,0)|^{2}$ and the other for the idler beam
with the amplification rate $|V_{s}(-Q,0)|^{2}$.\cite{gigi2} According to
Eqs. (\ref{40}), these two patterns are the same as the ordinary fringe (see
Eq. (\ref{8})) and can be identified by the polarization and separated in
space when the input beam is well tilted in incidence. However, for type I
crystal, the one-photon interference pattern is written as%
\begin{eqnarray}
|W(x,Q)|^{2} &=&\frac{kA^{2}}{f}\{|U(Q,0)|^{2}\widetilde{T}^{2}(\frac{kx}{f}%
-Q)+|V(-Q,0)|^{2}\widetilde{T}^{2}(\frac{kx}{f}+Q)  \notag \\
&&+[U(Q,0)V^{\ast }(-Q,0)\widetilde{T}(\frac{kx}{f}-Q)\widetilde{T}(\frac{kx%
}{f}+Q)+\text{c.c.}]\}.  \label{42}
\end{eqnarray}%
The first and second terms illustrate the two interference patterns created
by two stimulated beams, i.e. the signal beam with the transverse wavevector 
$Q$ and the idler beam with the transverse wavevector $-Q$. The third term
shows an additional coherent superposition of the two interferences of the
converted beams. In result, the interference pattern can be different from
the ordinary one due to the "interference term" in the square brackets of
Eq. (\ref{42}). Only for the normal incidence, $Q=0$, the interference
pattern is the same as the ordinary one. Figures 5a and 5b show the density
patterns of the one-photon interference by varying the transverse wavevector
of the input beam for type I and type II crystals, respectively. In Fig. 5a,
the fringes alternate between onset and offset when the transverse
wavevector $Q$ of the input field is increased until the two fringes are
well apart. The offset is due to the destructive interference of two
indistinguishable stimulated beams when they are folded. However, for type
II crystal, we put two interference patterns for the signal and idler beams
together in Fig. 5b, in which the right part is for the signal field, and is
stronger than the left part (idler). This corresponds to the measurement
when the detection is insensitive to the polarizations. Different from type
I crystal, when the transverse wavevector $Q$ of the input field is
increased, the bright and dark spots of the patterns exchange alternatively
until the two fringes are apart. This feature comes from the incoherent
superposition of the two patterns for the signal and idler beams.

We go through a long derivation, using the unitary transformation (\ref{11})
and the bosonic commutation relation, and obtain the second-order
correlation function 
\begin{align}
G^{(2)}(x_{1},x_{2})& =[\left\vert W_{i}(x_{1},Q)\right\vert
^{2}+M_{i}(x_{1},x_{1})][\left\vert W_{s}(x_{2},Q)\right\vert
^{2}+M_{s}(x_{2},x_{2})]  \label{43} \\
& +[W_{i}(x_{1},Q)W_{s}(x_{2},Q)N_{is}^{\ast }(x_{1},x_{2})+\text{c.c.}%
]+\left\vert N_{is}(x_{1},x_{2})\right\vert ^{2}  \notag \\
& +\delta _{is}\{[W^{\star }(x_{1},Q)W(x_{2},Q)M(x_{1},x_{2})+\text{c.c.}%
]+\left\vert M(x_{1},x_{2})\right\vert ^{2}\}.  \notag
\end{align}%
Again, Eq. (\ref{43}) can describe two types of crystals. For type I
crystal, the subscripts $i$ and $s$ are omitted and $\delta _{is}=1$,
whereas for type II crystal $\delta _{is}=0$. Similar to Eq. (\ref{g2}), the
first term is the product of the two one-photon interferences for the
converted beams. The second and the fourth terms exhibit the coupling of the
interferences between the stimulated and spontaneous processes. This result
is similar to the discussion of the image amplification in the optical
parametric amplification in which $W(x)$ describes an amplified image.\cite%
{wkg} We note that the interferences coming from the stimulated process are
uncorrected in positions. This reflects that the stimulated process does not
includes the entanglement in transverse wavevectors for the two converted
beams. Therefore, the quantum lithography cannot be achieved in the
stimulated parametric amplification by input a coherent beam. In Figs. 6, we
plot the density patterns of the two-photon interferences by varying the
input direction of the signal field. Figures 6a and 6b show patterns of $%
G^{(2)}(x,x)$ and $G^{(2)}(x,-x)$ for type I crystal, respectively. These
two patterns are similar to the one-photon interference case with the
exception of that the sub-wavelength fringe coming from the spontaneous
process appears in the central part. In Fig. 6a, the fringe of the signal
part (the right side) is stronger than the idler part (the left side),
whereas in Fig. 6b, the pattern is mirror-symmetry since exchange of two
detectors at a pair of symmetric positions makes no difference. However,
Figs. 6c and 6d show patterns of $G^{(2)}(x,x)$ and $G^{(2)}(x,-x)$ for type
II crystal, respectively. Correspondingly, Fig. 6c is similar to the
one-photon interference case with the exception of that of the spontaneous
contribution. In Fig. 6d, we define $x$ as the transverse position for the
detector sensitive to the polarization of the signal beam, so that it is
comparable to the one-photon case for the signal beam (the right part in
Fig. 5b). Furthermore, in Fig. 6d, the spontaneous process contributes a
homogeneous background instead of a sub-wavelength fringe in Fig. 6c. If two
detectors in the joint-intensity measurement are polarization-insensitive,
one observes a symmetric patterns which is the mirror-symmetry of Fig. 6d.

\section{Conclusions}

In summary, we formulate the first- and second-order correlation functions
in the Young's double-slit interference for both spontaneous and stimulated
parametric down-conversions. The results reveal the relations between the
first- and second-order correlations, and hence it can explain the
complementarity of coherence and entanglement. We show that the nonclassical
sub-wavelength two-photon interference can occur macroscopically in a
general spontaneous parametric process. For a high gain of SPDC, in which
the converted beams contain a huge number of photons, the sub-wavelength
interference pattern is intensive with a substantial visibility. This makes
the quantum lithography technology in practicability. Moreover, we find an
alternative way to observe the sub-wavelength interference for only type I
crystal, in which a joint-intensity measurement is performed by a pair of
one-photon detectors placed at the symmetric positions. The advantage of
this method is that the quantum lithography in type I of SPDC can be
performed by two one-photon detectors instead of a two-photon detector which
could be unavailable. Since, this effect occurs only in a higher gain of
SPDC, it reflects macroscopically quantum nature for the entangled beams
containing a huge number of photons. The two ways of observations compete
for the visibility, so that at the low gain of SPDC of type I the
interference of the first observation reaches perfectness while the second
observation disappears. In the stimulated process, the one-photon and
two-photon interference patterns generated by a pair of stimulated
down-converted beams are alike. For type I crystal, since two converted
beams are polarization-indistinguishable, the coherent superposition of two
converted beams causes a secondary interference which may fade the fringe
when two beams are not well apart. However, this effect does not exist for
type II crystal because of the distinguishability in polarization for the
two stimulated beams.

\section{Acknowledgment}

This research was supported by the National Program of Fundamental Research
No. 2001CB309310 and the National Natural Science Foundation of China,
Project Nos. 10074008, 60278021 and 10174007.

Captions of Figures :

Fig. 1. Schemes of Young's double-slit interference with a\ convex lens: (a)
a one-photon (two-photon) detector measures one-photon (two-photon)
intensity distribution; two one-photon detectors measure joint-intensity
distribution at a pair of symmetric positions. (b) for type II crystal,
two-photon intensity distribution is measured by a polarizing beamsplitter
(PBS)\ and two one-photon detectors.

Fig. 2. One-photon (solid line) and two-photon (dashed line) double-slit
interference patterns for a coherent beam.

Fig. 3. Visibilities versus the gain of SPDC for different collinear
phase-mismatching (a) $\Delta _{0}=-5.85$;\ (b) $\Delta _{0}=0$; and\ (c) $%
\Delta _{0}=5.85$. Solid and Dashed lines designate $\mathcal{V}_{1}$ for
type II and type I crystals, respectively; Dotted line designates $\mathcal{V%
}_{2}$ for type I crystal.

Fig. 4. Two-photon interference patterns versus the normalized bandwidth of
SPDC $q_{0}b/(2\pi )$: (a) $G^{(2)}(X,X)$ for type I crystal; (b) $%
G^{(2)}(X,-X)$ for type I crystal; (c) $G^{(2)}(X,X)$ for type II crystal;
and (d) $G^{(2)}(X,-X)$ for type II crystal, where $X=xkb/(2\pi f)$ is the
normalized transverse position in the detection plane. The gains are set as $%
g=(1/2)\log 1.5$ in (a) and (c), and $g=(1/2)\log 10$ in (b) and (d). The
other parameters in Figs. 4-6 are taken as the phase matching $\Delta _{0}=0$%
, the ratio $b/d=0.2$, and $\frac{kb\Omega _{0}}{4\pi ^{2}f}=1$ for an
arbitrary unit of the correlation functions.

Fig. 5. Stimulated one-photon interference patterns by varying the
normalized transverse wavevector of the input field $Qb/(2\pi )$ for (a)
type I crystal and (b) type II crystal. The normalized bandwidth and the
input intensity are taken as $q_{0}b/(2\pi )=2$ and $\frac{2kb^{2}}{\pi f}%
A^{2}=1$, respectively.

Fig. 6. Stimulated two-photon interference patterns by varying the
normalized transverse wavevector of the input field $Qb/(2\pi )$: (a) $%
G^{(2)}(x,x)$ for type I crystal; (b) $G^{(2)}(x,-x)$ for type I crystal;
(c) $G^{(2)}(x,x)$ for type II crystal; and (d) $G^{(2)}(x,-x)$ for type II
crystal. The normalized bandwidth and the input intensity are the same as in
Fig. 5.

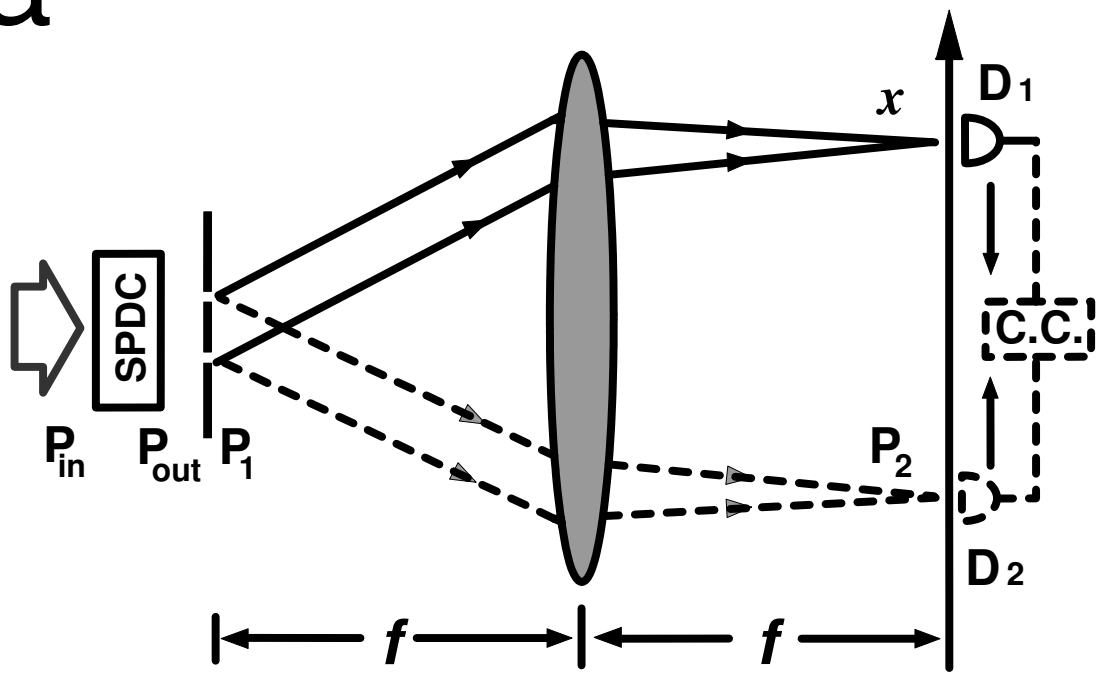
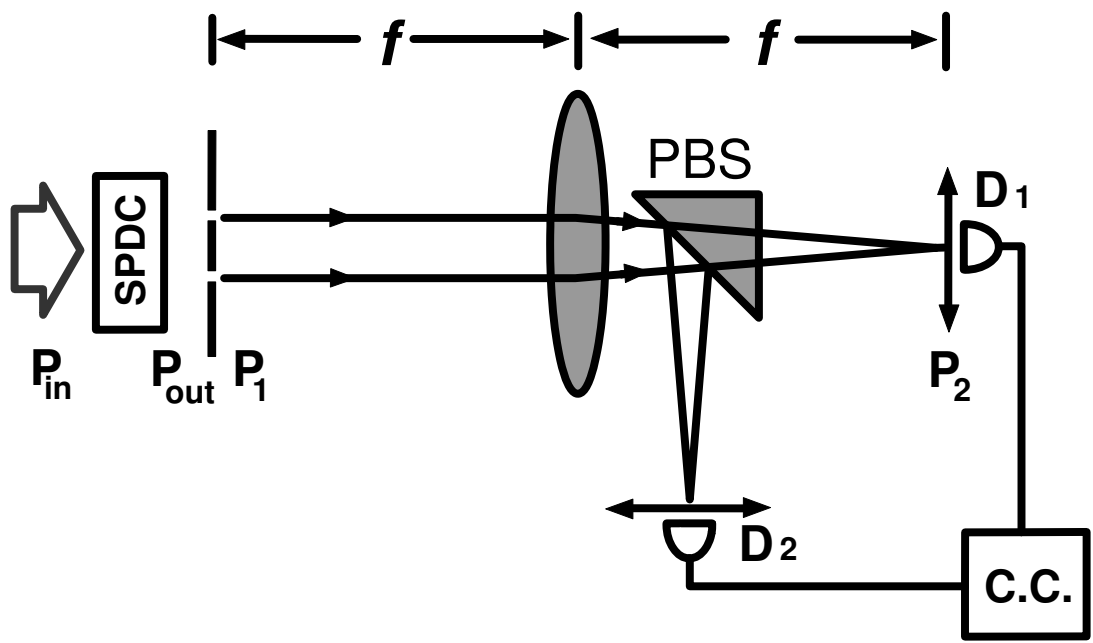
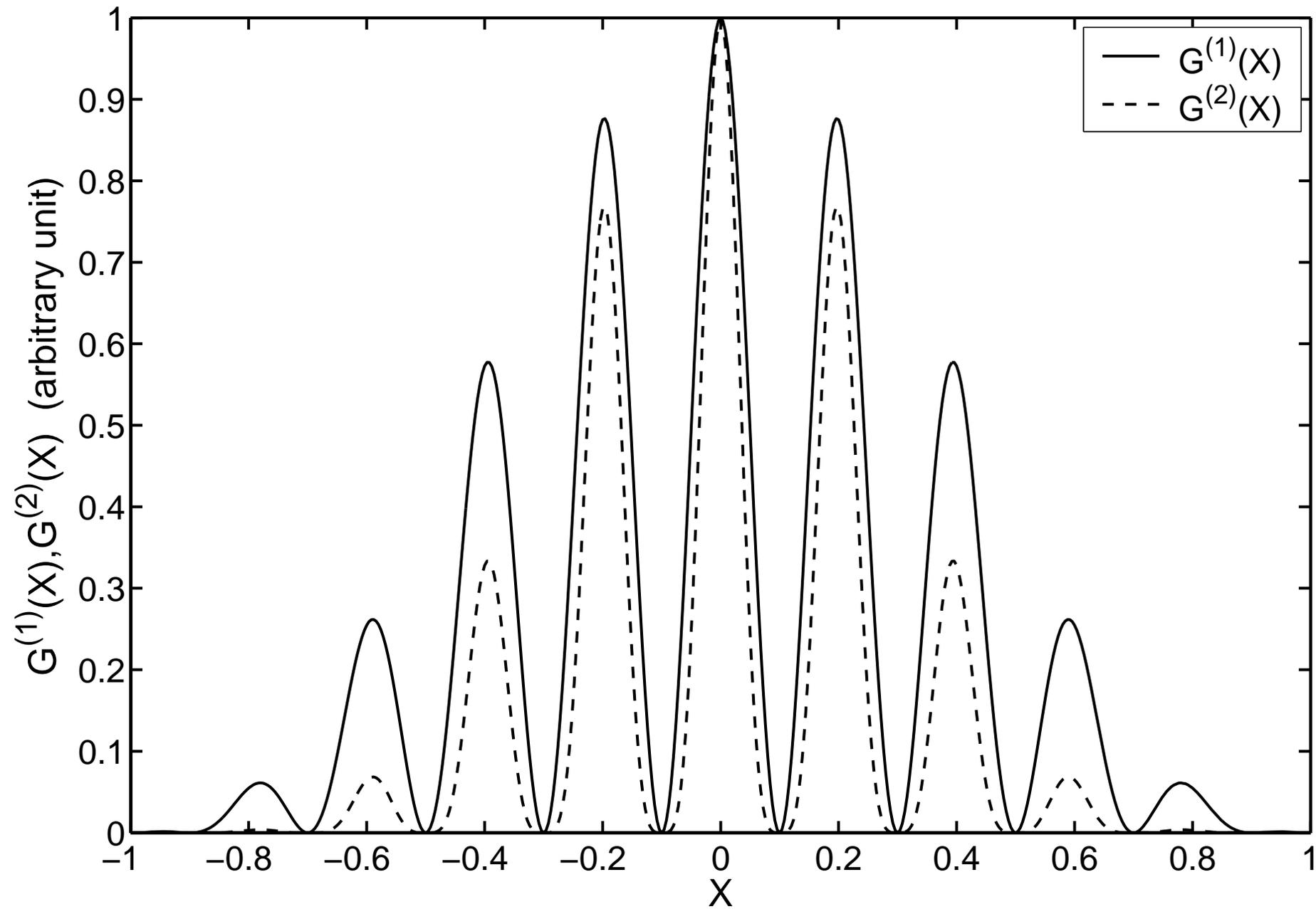
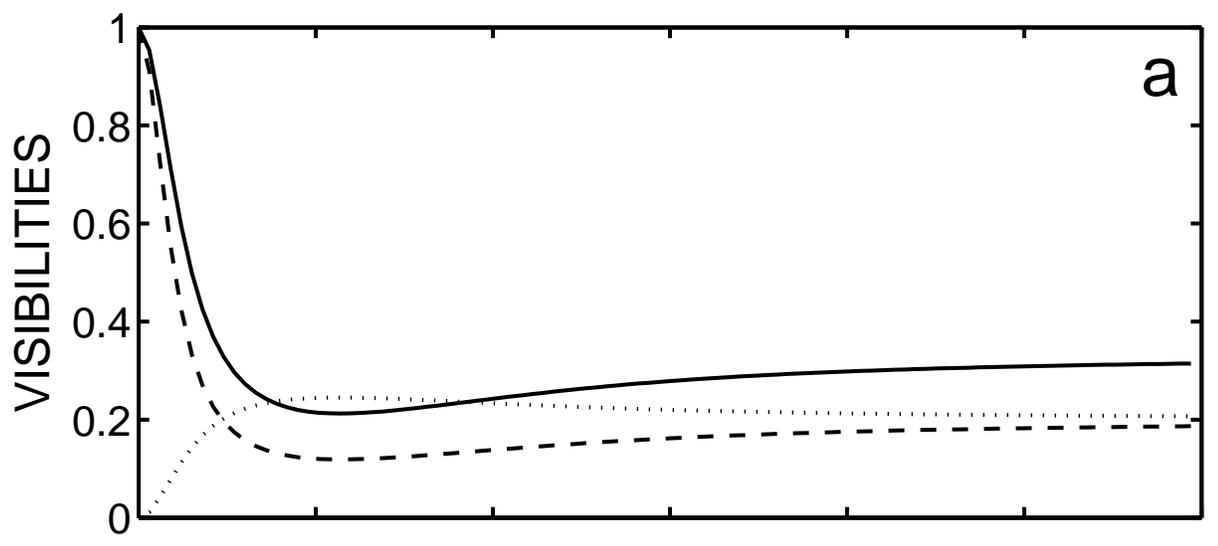
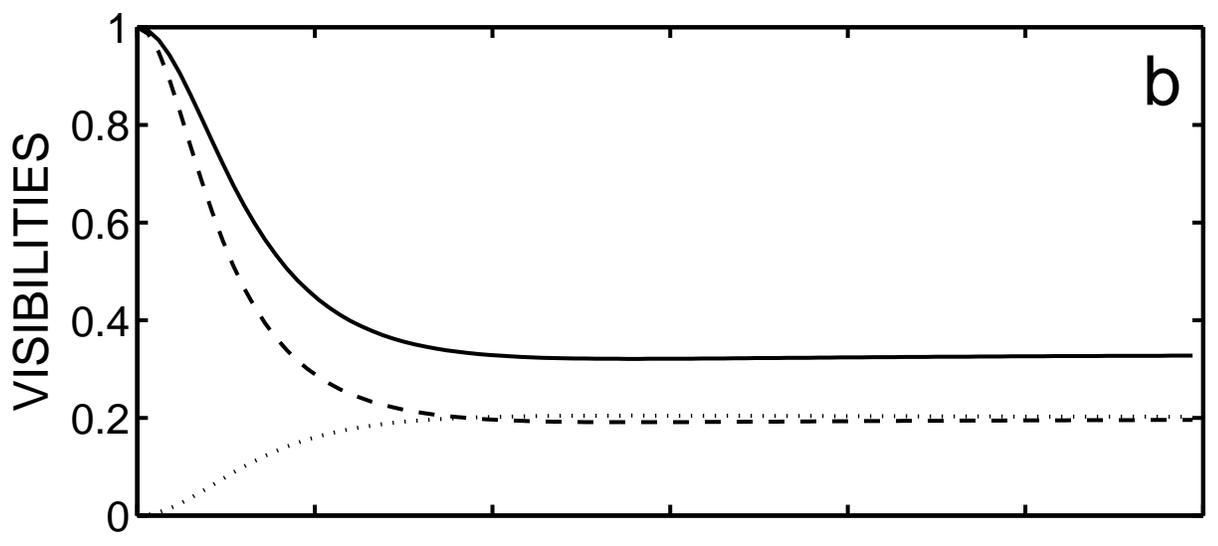
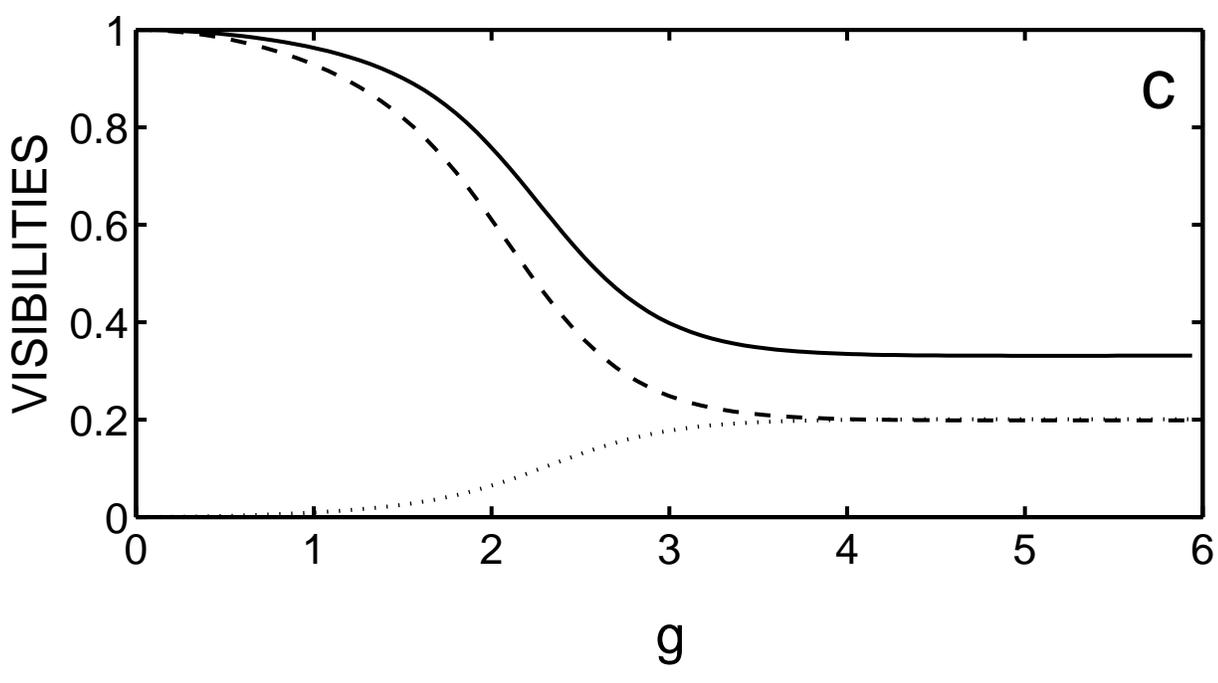
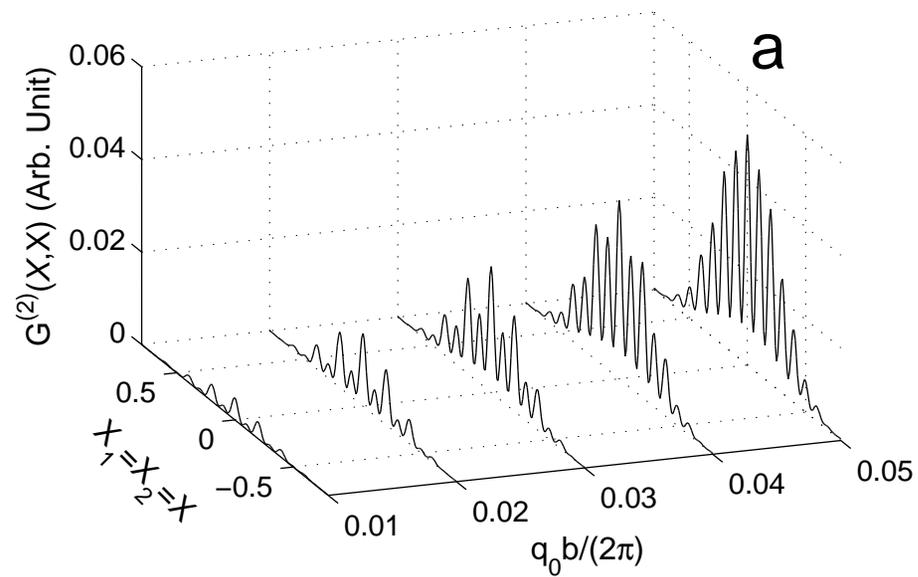
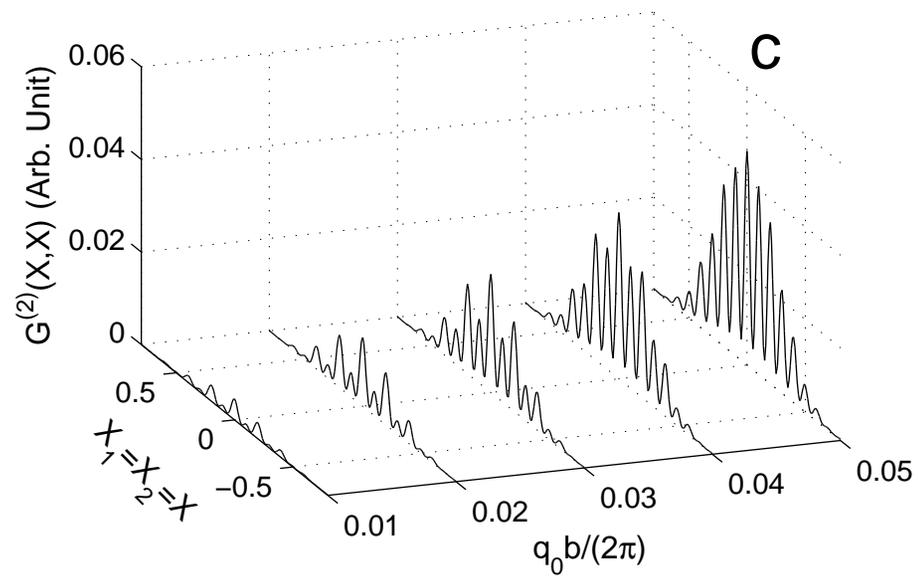
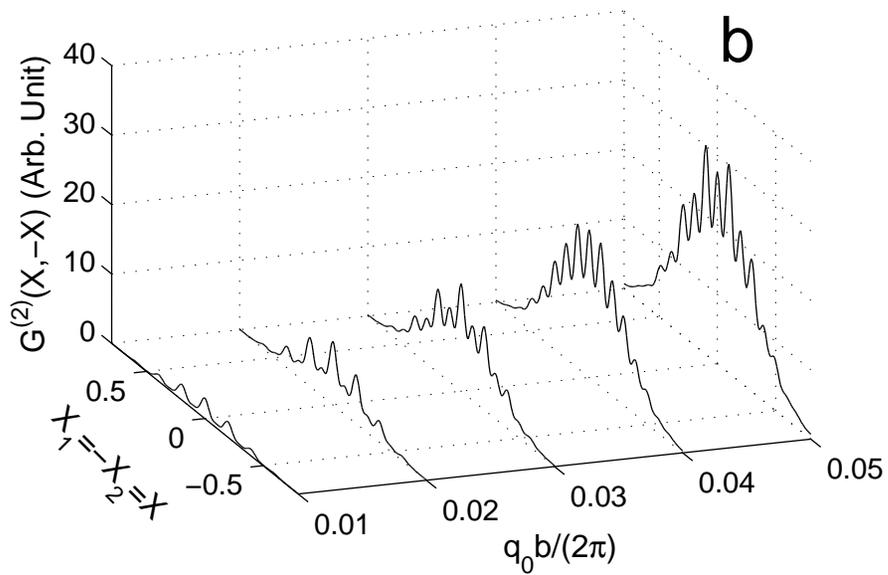
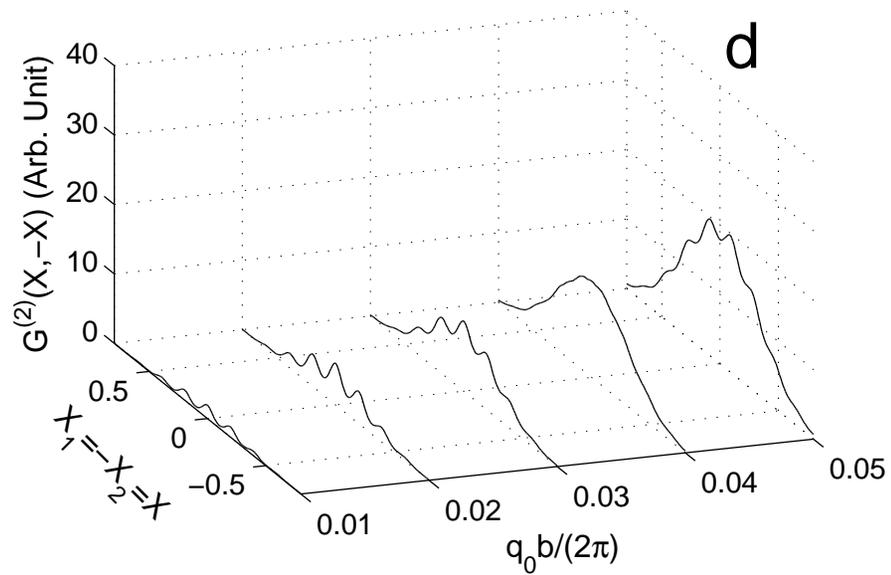
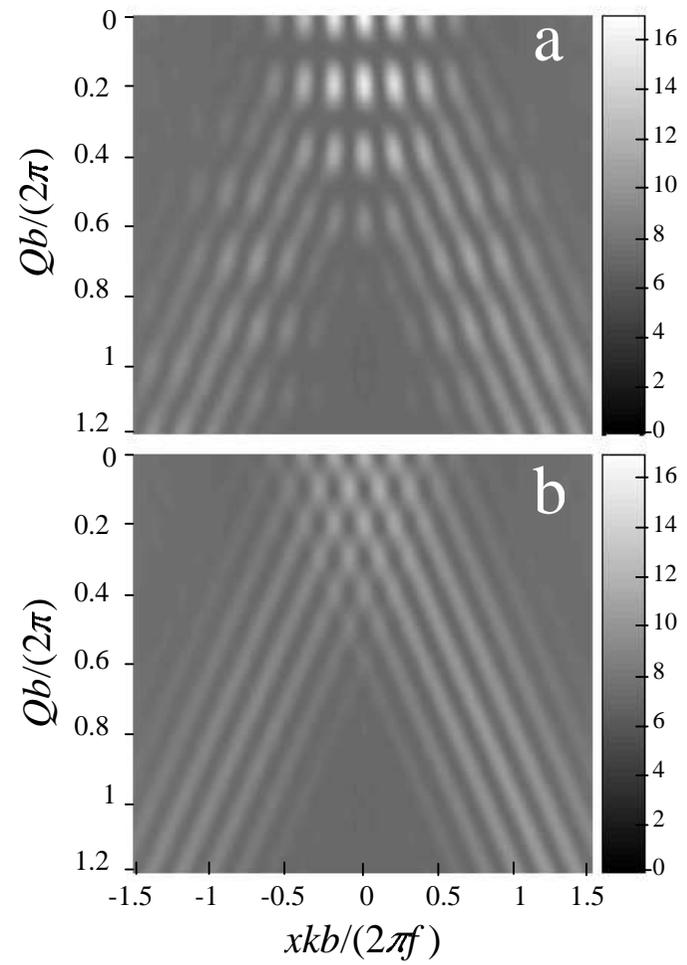
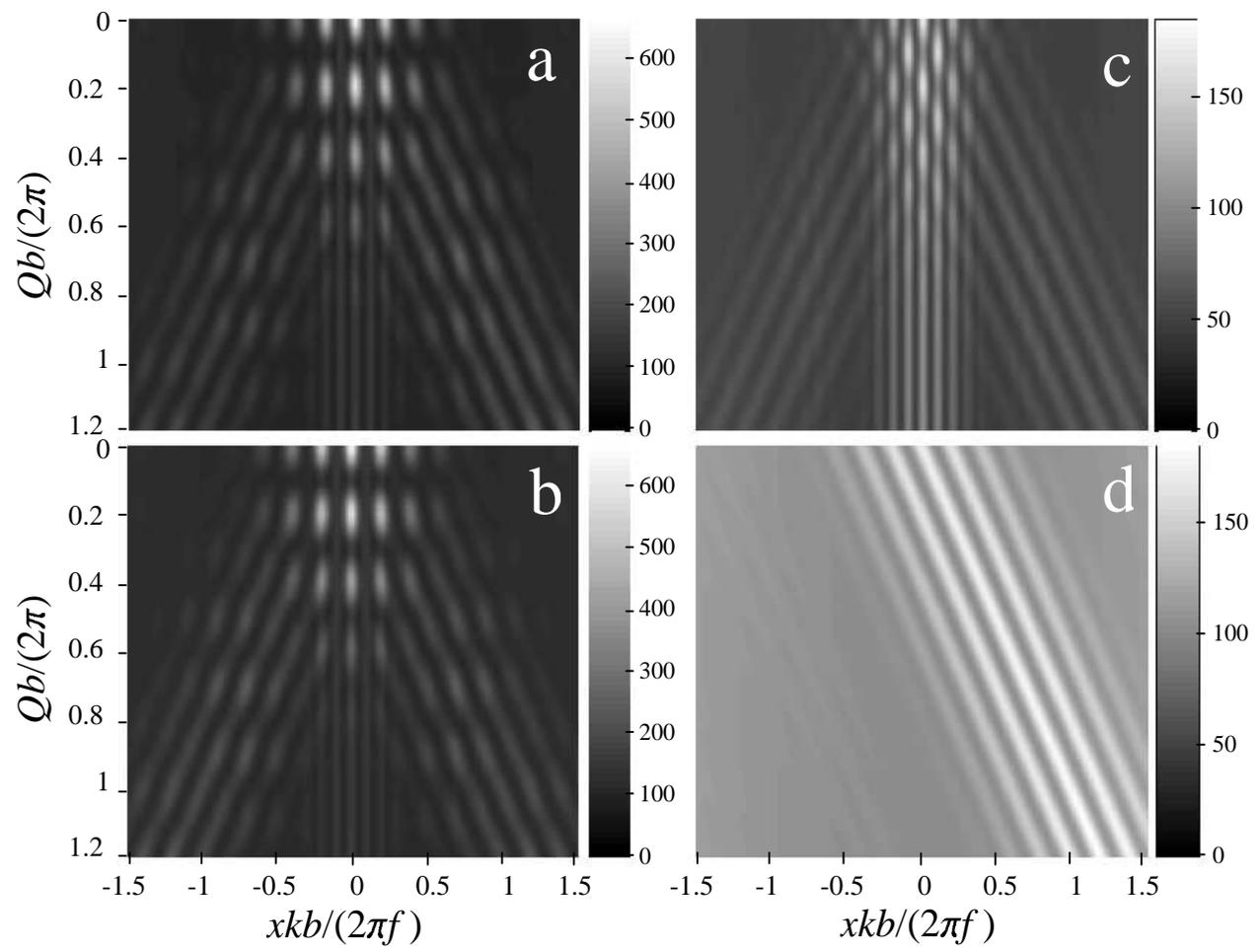
\end{document}